\documentclass[final]{aipproc}

\usepackage{slashed}

\layoutstyle{6x9}

\def\E{{\mathrm e}}

\def\be{\begin{equation}}
\def\ee{\end{equation}}
\def\bq{\begin{eqnarray}}
\def\eq{\end{eqnarray}}

\newcommand{\lsim}{\mathrel{\rlap{\lower4pt\hbox{\hskip1pt$\sim$}}
    \raise1pt\hbox{$<$}}}         
\newcommand{\gsim}{\mathrel{\rlap{\lower4pt\hbox{\hskip1pt$\sim$}}
    \raise1pt\hbox{$>$}}}         

\begin{document}

\title{Phenomenology of transverse-spin and transverse-momentum 
effects in hard processes}

\classification{13.88.+e, 12.38.-t}

\keywords      {transverse spin, transverse momentum, SIDIS, 
Drell-Yan, asymmetries}

\author{Vincenzo Barone}{
  address={Di.S.T.A., Universit{\`a} del Piemonte Orientale `A. Avogadro' \\
and INFN, Gruppo Collegato di Alessandria, 
15121 Alessandria, Italy}
}

\begin{abstract}
Some recent analyses of single-spin and azimuthal 
asymmetries in SIDIS and Drell-Yan processes, 
focusing in particular on Collins, Sivers and Boer-Mulders effects, 
 are briefly reviewed. The perspectives for future phenomenological 
studies are 
also outlined. 
 
\end{abstract}

\maketitle

\section{Introduction}

Transverse spin
and transverse momentum of quarks and/or hadrons
correlate with each other in various ways, 
giving rise to a number of transverse-momentum dependent 
distributions (TMD's), some of which are 
leading-twist quantities. TMD's
manifest themselves 
in single-spin and azimuthal asymmetries in 
partially inclusive hard processes  
\cite{Barone:2010a}.  
In the last decade, semiinclusive 
deep inelastic scattering (SIDIS) experiments
(HERMES, COMPASS, CLAS)  have investigated these 
observables and shown that they are non vanishing 
and relatively 
sizable.

\section{State of the art}

Most of the 
phenomenological work focuses on three distributions functions: 
the transversity distribution $h_1(x)$, 
 the Boer-Mulders function $h_1^{\perp} (x, k_T^2)$ (spin 
asymmetry of transversely polarized quarks inside an unpolarized target) 
and the Sivers function $f_{1T}^{\perp} (x, k_T^2)$ 
(azimuthal asymmetry of unpolarized quarks inside a transversely 
polarized nucleon). 
The first two combine in SIDIS 
with the Collins fragmentation function $H_1^{\perp}$, which describes 
the fragmentation of transversely polarized quarks into an unpolarized 
hadron.   
The processes considered by present phenomenological 
analyses are:              $e \, p^{\uparrow} \, \rightarrow 
               \, e' \, \pi \, X$ 
(Collins and Sivers effects with 
                different angular distributions), 
                   $e \, p \, \rightarrow 
               \, e' \, \pi \, X$,  
                  $p \, p \, 
            \rightarrow \, \mu^+ \, \mu^- \, X$
                (Boer-Mulders effect), 
$ e^+ \, e^- \, \rightarrow \, \pi \, \pi \, X$ (Collins effect).

The parton-model expressions  
of the SIDIS structure functions involving the 
three TMD's mentioned above are 
\bq
{\rm Collins} & & 
F_{UT}^{\sin (\phi_h + \phi_S)} = 
{\mathcal C} \left [ - \frac{\hat{h} \cdot \vec \kappa_T}{M_h} \, 
h_1 H_1^{\perp} \right ] 
 \\
{\rm Sivers} & &   
F_{UT, T}^{\sin (\phi_h - \phi_S)} = 
{\mathcal C} \left [ - \frac{\hat{h} \cdot \vec k_T}{M} \, 
f_{1T}^{\perp} D_1 \right ]
 \\ 
{\rm Boer-Mulders} & &  
F_{UU}^{\cos 2 \phi_h} = 
{\mathcal C} 
\left [ -\frac{2 (\hat{h} \cdot \vec k_T) (\hat{h} 
\cdot \vec \kappa_T) - \vec k_T \cdot \vec \kappa_T}{M M_h} 
\, h_1^{\perp} \, H_1^{\perp} \right ] 
\eq
where $\vec k_T$ ($\vec \kappa_T$) 
is the transverse momentum of the incoming (fragmenting) quark, 
the apices indicate the azimuthal modulation ($\phi_h$ 
and $\phi_S$ being the azimuthal angles of the 
final hadron and of the target spin, respectively) and ${\mathcal C}$ 
is a convolution in the transverse momentum space.

TMD's are usually written 
as factorized functions  of $x$ and $k_T$, and their 
transverse-momentum dependence is often assumed to have a Gaussian 
form.  
A typical parametrization for the Sivers 
and the Boer-Mulders functions is 
\be
 f_{1T}^{\perp}(x, k_T^2), \;   
 h_{1}^{\perp} (x, k_T^2)   
\sim x^{\alpha} (1 - x)^{\beta} \, \E^{- k_T^2/\langle 
k_T^2 \rangle} \, f_1(x)\,,  
\label{param}
\ee
where $f_1(x)$ is the ordinary number density. 
Due to the kinematics of current experiments and to 
the structure of SIDIS observables,  
high-$x$ tails and antiquark distributions 
are at present largely 
unconstrained.

A combined analysis of the SIDIS data on the 
Collins asymmetry from HERMES and COMPASS, and 
of the $e^+ e^-$ Belle data, was performed by Anselmino et al. 
\cite{Anselmino:2007fs,Anselmino:2008jk} 
and led to the first extraction of the $u$ and 
$d$ transversity distributions, which  
turned out to have opposite signs, with $\vert h_{1}^d \vert$ 
smaller than $\vert h_{1}^u \vert$.

The Sivers  asymmetry has been 
measured by HERMES and COMPASS and phenomenologically studied 
in \cite{Anselmino:2008sga} and in 
\cite{Arnold:2008ap}. The resulting Sivers functions 
for $u$ and $d$ have comparable magnitudes 
and opposite signs  
($f_{1T}^{\perp u} <0$, $f_{1T}^{\perp d} >0$).

\begin{figure}
\includegraphics[width=0.38\textwidth,angle=-90]
{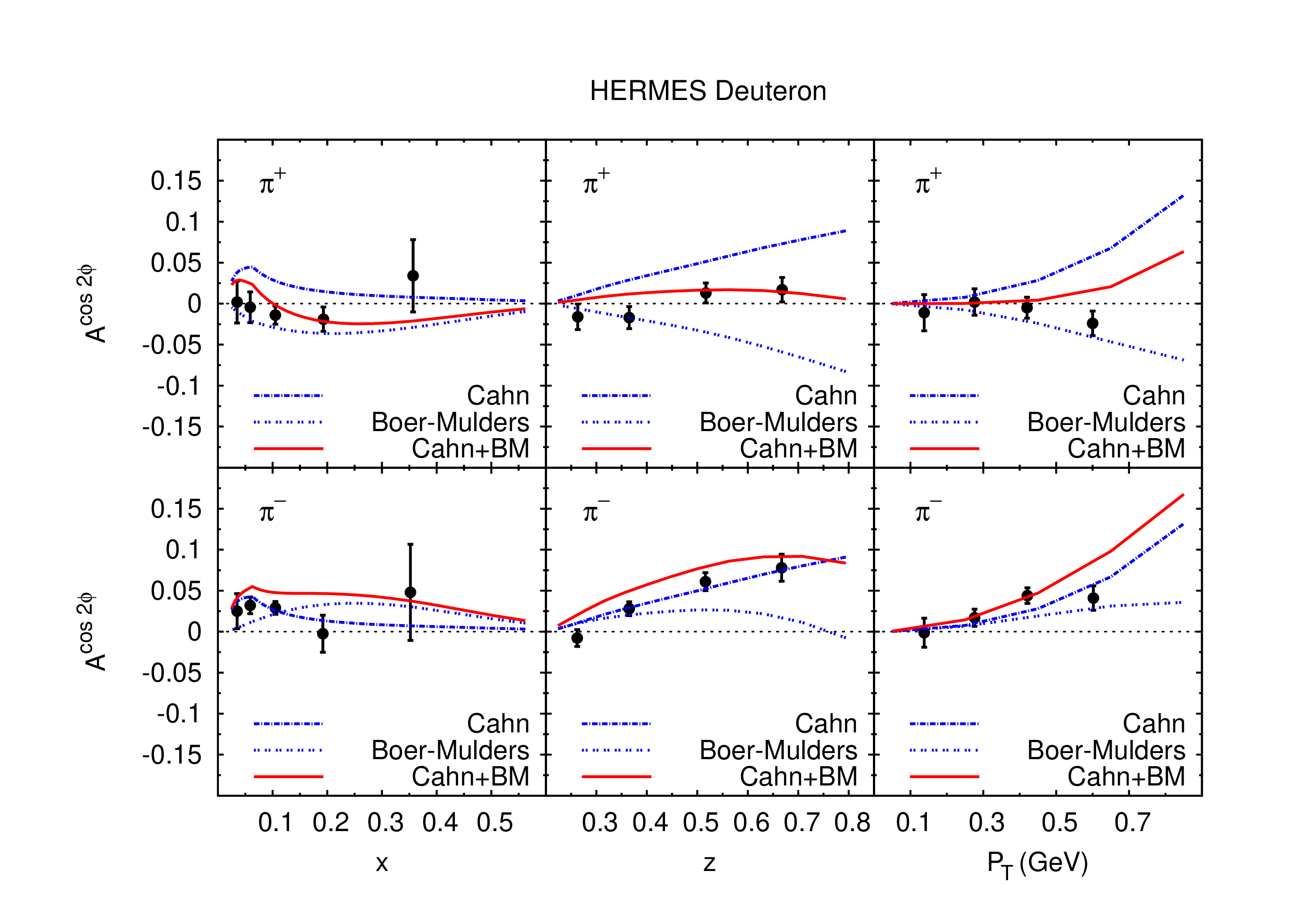}
\hspace{-0.5cm} 
\includegraphics[width=0.38\textwidth,angle=-90]
{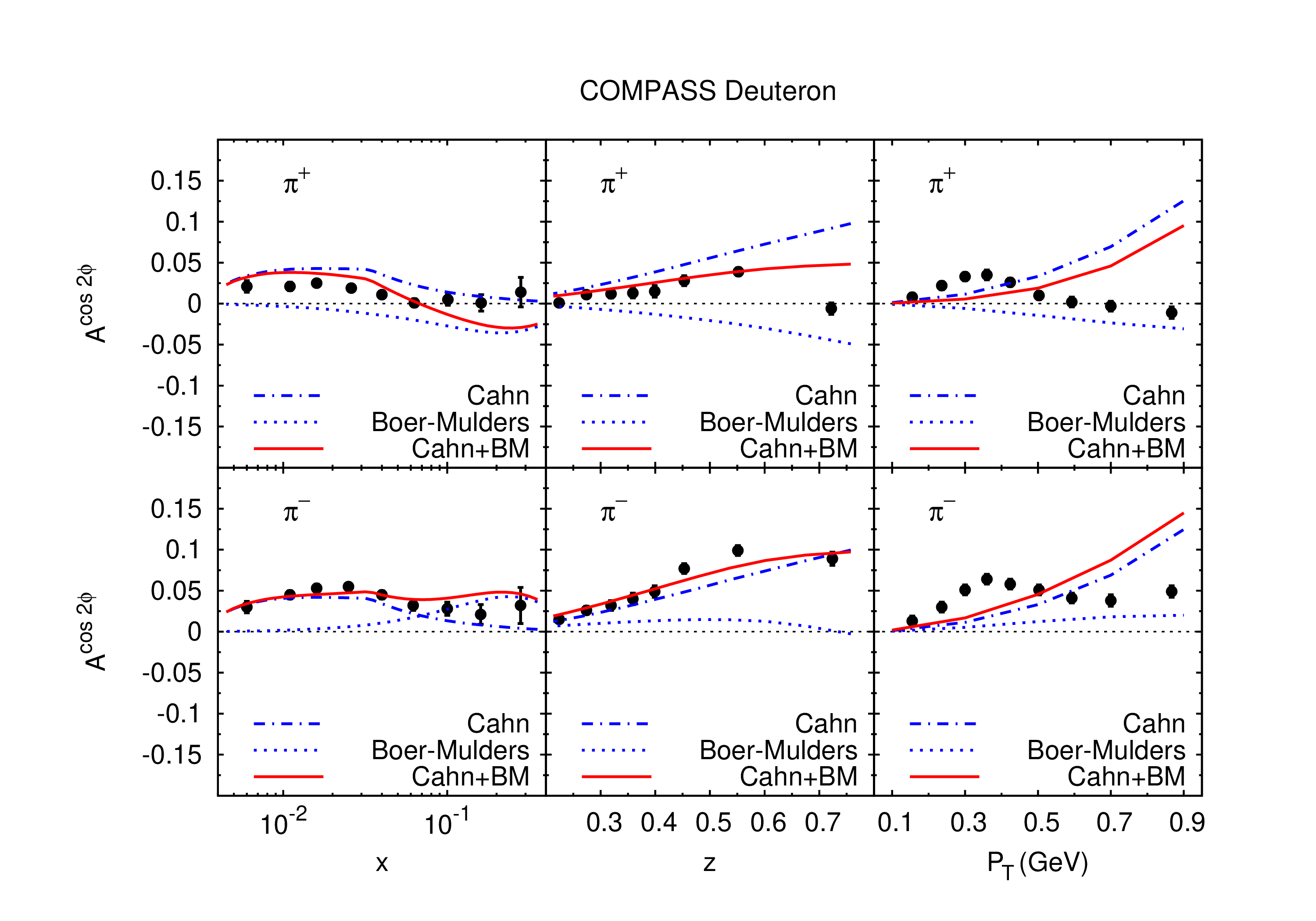}
\caption{The preliminary results for the
$\cos 2\phi$ spin-independent azimuthal asymmetries for deuteron
from HERMES \cite{Giordano:2009hi} (left) and COMPASS 
\cite{Kafer:2008ud} (right) as functions of 
$x$, $z$ and $P_{h \perp}$ compared with
the fit of~\cite{Barone:2009hw}.}
\label{fig_costwophi}
\end{figure}

The $\cos 2 \phi_h$ asymmetry in unpolarized SIDIS 
at small transverse momentum 
provides information on the Boer-Mulders function.  
In \cite{Barone:2008tn} it was predicted 
that the $\pi^-$ asymmetry should be larger than the $\pi^+$ 
asymmetry, as a consequence of the Boer-Mulders effect. 
This prediction has been substantially confirmed by 
the experimental results. 
A fit to the HERMES \cite{Giordano:2009hi} 
and COMPASS \cite{Kafer:2008ud} preliminary data has been 
performed in \cite{Barone:2009hw}. 
It assumes that $A_{UU}^{\cos 2 \phi_h}$ can be 
described by the leading-twist Boer-Mulders component 
and by the so-called Cahn term \cite{Cahn:1989yf}
\be
F_{UU, {\rm Cahn}}^{\cos 2 \phi_h} = 
\frac{M^2}{Q^2} \, 
{\mathcal C} 
\left [ \frac{(2 (\hat{h} \cdot \vec k_T)^2 -  k_T^2)}{M^2} 
 f_1 \, D_1 \right ]
\ee
which is however only part 
of the full twist-4 contribution, still unknown. 
As the available data do not allow a complete determination  
of the $x$ and $k_T$ dependence of $h_1^{\perp}$,  
the Boer-Mulders functions are simply taken to be proportional to 
the Sivers functions of \cite{Anselmino:2008sga}, $h_1^{\perp q} 
= \lambda_q f_{1T}^{\perp q}$, and the parameters $\lambda_q$ 
are obtained from the fit.  
The result, $h_{1}^{\perp u} \simeq 2 \, f_{1T}^{\perp u}$,  
$h_1^{\perp d} \simeq -  \, f_{1T}^{\perp d}$, is  
consistent with expectations 
from the impact-parameter picture 
and lattice QCD. 
The comparison with the data 
is shown in Fig.~\ref{fig_costwophi}.

\begin{figure}
\includegraphics[width=0.25\textwidth,angle=-90]
{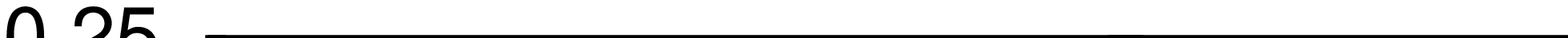} 
\caption{The $\nu$ parameter in $pD$ Drell-Yan production: the 
fit of \cite{Barone:2010gk} vs. the E866/NuSea data \cite{Zhu:2006gx}.} 
\label{fig_dy}
\end{figure}

The $\cos 2 \phi$ asymmetry has been measured also 
in  unpolarized Drell-Yan (DY) production, where it  
is represented by the so-called $\nu$ parameter. At small $Q_T$ 
this quantity is dominated by the 
Boer-Mulders contribution and is given by 
\be
 \nu = \frac{2 \,  W_{UU}^{\cos 2 \phi}}{W_{UU}^1 + W_{UU}^2} 
\,, \;\;\;\;\;
 W_{UU}^{\cos 2 \phi} = \frac{1}{3} \, {\mathcal C} \left [ 
\frac{2 (\hat{h} \cdot \vec k_{1T}) 
(\hat{h} \cdot \vec k_{2T}) - \vec k_{1T} 
\cdot \vec k_{2T}}{M_1 M_2} \, h_1^{\perp} \bar h_1^{\perp} 
\right ].
\ee
The E866/NuSea Collaboration at 
FNAL has presented results for the $\nu$ asymmetry 
in $pD$ \cite{Zhu:2006gx} and $pp$ \cite{Zhu:2008sj}
collisions, from which one 
can get some information on the antiquark 
Boer-Mulders distributions. The analysis of the 
SIDIS $\cos 2\phi$  
distribution  has been extended to 
the corresponding DY observable in \cite{Barone:2010gk}. 
The quality of the 
fit is shown in Fig.~\ref{fig_dy}. The combined SIDIS and DY 
analysis allows 
determining both the magnitude and the sign of 
the quark and antiquark distributions.

\section{Perspectives}

In spite of its important achievements, 
the phenomenology of TMD's is still in its infancy. The mature 
stage will be represented by truly global analyses. 
 These should: 
{\it i)} incorporate the 
exact evolution of TMD's;    
{\it ii)} take all 
perturbative and non-perturbative effects into account 
and 
fit simultaneously polarized and unpolarized cross sections;   
{\it iii)} use datasets with larger statistics and  
wider kinematics. 
More SIDIS data are expected from JLab and from 
future facilities (EIC), but in the 
short-to-medium term the main improvement will 
come from  polarized DY measurements: 
COMPASS ($\pi^{\pm}p^{\uparrow}$), 
PANDA ($\bar p p^{\uparrow}$), 
PAX ($\bar p^{\uparrow} p^{\uparrow}$), 
J-PARC, NICA, RHIC ($p^{\uparrow} p$). 
A list of DY observables 
is ($\phi$ is the azimuthal angle of dileptons in 
the Collins-Soper frame)
\begin{itemize} 
\item
$ \pi^{\pm} p:
\hspace{1cm} 
 \bar h_1^{\perp \pi}  \otimes 
h_1^{\perp p}\; \cos 2 \phi$ 

Driven by valence. 
Expected to be large (10-15\%), but involves the 
Boer-Mulders function of the pion. 

\item  
$ \pi^{\pm} p^{\uparrow}: 
\hspace{1cm} 
 \bar f_1^{\pi}  \otimes 
f_{1T}^{\perp p} \; \sin (\phi - \phi_S) 
+  \bar h_1^{\perp \pi}  \otimes 
h_1^p \; \sin (\phi + \phi_S)$

Driven by valence. The  $\sin (\phi - \phi_S)$ asymmetry 
probes the Sivers function of the proton
(the unpolarized distribution of the pion is fairly well 
known). 

\item
$ pp: 
\hspace{1cm} 
 \bar h_1^{\perp p}  \otimes h_1^{\perp p} \; 
\cos 2 \phi
$

Involves the sea. Known to be small (few percent).

\item
$p p^{\uparrow}: 
\hspace{1cm} 
 \bar f_1^{p} \otimes 
f_{1T}^{\perp p}  \; \sin (\phi - \phi_S)+ 
\bar h_1^{\perp p}  \otimes 
h_1^p \; \sin (\phi + \phi_S)
$

Involves the sea,  but is useful to extract the Sivers function.

\item
$\bar p p: 
\hspace{1cm} 
 h_1^{\perp p}  \otimes h_1^{\perp p} \; \cos 2 \phi 
$

Driven by valence. Expected to be large. 
Ideal to extract the Boer-Mulders function. 

\item
$\bar p p^{\uparrow}: 
\hspace{1cm} 
f_1^p \otimes 
f_{1T}^{\perp p}  \; \sin (\phi - \phi_S) 
+  h_1^{\perp p}  \otimes 
h_1^p \; \sin (\phi + \phi_S)
$

Driven by valence. The $\sin (\phi - \phi_S)$ 
asymmetry  ideal to extract the Sivers function.

\item
$\bar p^{\uparrow} p^{\uparrow}: 
\hspace{1cm} 
h_1^p \otimes h_1^p \, \cos (2 \phi - \phi_{S_1} - \phi_{S_2})
$

Driven by valence. Ideal to extract the transversity. 

\end{itemize} 

As one can see, 
polarized DY 
processes probe various 
combinations of TMD's and promise to become a fundamental 
ingredient of future phenomenological analyses.

\begin{theacknowledgments}
I  thank A.~Prokudin, S.~Melis, 
F.~Bradamante, A.~Martin, M.~Anselmino 
and E.~Boglione 
for many useful discussions. Work supported 
in part by MIUR (PRIN 2008).

\end{theacknowledgments}

\end{document}